\documentclass[structabstract]{aa}  
\usepackage{graphicx}
\usepackage{txfonts}
\usepackage{natbib} 
\bibpunct{(}{)}{;}{a}{}{,} %
\begin{document}
   \title{Galaxy Luminosity Functions in WINGS clusters\thanks{Based on observations collected at the European Organisation for Astronomical Research in the Southern Hemisphere, Chile. Progs. ID  67.A-0030, 68.A-0139, and 69.A-0119}}

   \author{A. Moretti
          \inst{1,2}
          \and
          D. Bettoni \inst{1}
          \and 
          B. Poggianti \inst{1}
          \and 
          G. Fasano \inst{1}
          \and
          J. Varela \inst{3}
          \and
          M. D'Onofrio \inst{2}
          \and
          B. Vulcani \inst{4}
          \and
          A. Cava \inst{5}
          \and
          J. Fritz \inst{6}
          \and
          W. J. Couch \inst{7}
         \and
          M. Moles \inst{3}
          \and
          P. Kj\ae rgaard \inst{8}
          }

   \institute{INAF - Osservatorio Astronomico di Padova,
              Vicolo dell'Osservatorio, 5, Padova\\
              \email{alessia.moretti@oapd.inaf.it}
         \and
             University of Padova, Department of Physics and Astronomy,
             Vicolo dell'Osservatorio, 2, Padova
             \and
             Centro de Estudios de Fisica del Cosmos de Aragon, 
             Plaza de San Juan 1, 44001 Teruel, Spain
             \and
             Kavli Institute for the Physics and Mathematics of the Universe (WPI), Todai Institutes for Advanced Study, the University of Tokyo
             \and
             Observatoire de Gen\'eve, Universit\'e de Gen\'eve
             \and 
             Centro de Radioastronoma y Astrofsica, UNAM, Campus Morelia, A.P. 3-72, C.P. 58089, Mexico
             \and 
             Australian Astronomical Observatory, PO Box 915, North Ryde, NSW 1670 Australia
             \and
             Niels Bohr Institute, Juliane Maries Vej 30, 2100 Copenhagen, Denmark\\
                          }

   \date{Received ; accepted }

 
  \abstract
   {}
   {Using V band photometry of the WINGS survey, we derive galaxy luminosity functions (LF) in nearby clusters. This sample is complete down to $M_V=-15.15$, and it is homogeneous, thus allowing the study of an unbiased sample of clusters with different characteristics. }
   {We constructed the photometric LF for 72 out of the original 76 WINGS clusters, excluding only those without a velocity dispersion estimate. For each cluster we obtained the LF for galaxies in a region of radius=$0.5 \times r_{200}$, and fitted them with single and double Schechter's functions. We also derive the composite LF for the entire sample, and those pertaining to different morphological classes. Finally we derive the spectroscopic cumulative LF for 2009 galaxies that are cluster members.}
   {The double Schechter fit parameters are neither correlated with the cluster velocity dispersion, nor with the X--ray luminosity. Our median values of the Schechter's fit slope are, on average, in agreement with measurements of nearby clusters, but are less steep that those derived from large surveys, such as the SDSS.  Early--type galaxies outnumber late--types at all magnitudes, but both early and late types contribute equally to the faint end of the LF. Finally, the spectroscopic LF is in excellent agreement with the ones derived for A2199, A85 and Virgo, and with the photometric one at the bright magnitudes (where both are available).}
   {There is a large spread in the LF of different clusters. However, this spread is not caused by correlation of the LF shape with cluster characteristics such as X--ray luminosity or velocity dispersions.
   The faint end is flatter than what previously derived ($\alpha_f=-1.7$) at odds with what predicted from numerical simulations.}


  \maketitle
%

\section{Introduction}

Galaxy clusters are unique laboratories to study the environmental effects on galaxy evolution. How galaxies form and evolve can be studied using a variety of techniques, one of those being the galaxy luminosity function (LF). The LF, i.e. the number density of galaxies at a given luminosity, is one of the most fundamental statistics of galaxy populations. Its shape and variation with environment provide a crucial constraint on any model of galaxy evolution. 

The LF can be used as a diagnostic tool to search for changes in the galaxy population, for example the study of the shape of the LF with respect to the cluster-centric radius can give important insight into the dynamical processes working in clusters. Several studies show that quiescent and star--forming galaxies have very different LF \citep{madgwick02,Christlein03}. Galaxies in clusters have been often compared to galaxies in the field, at many different wavelengths, leading to results that are sometimes contradictory. \citet{DePropris98}, \citet{Christlein03}, \citet{Cortese03} and \citet{Bai06} found the cluster LF to be indistinguishable from field one, while other authors suggest that it has both brighter characteristic magnitudes and different faint end slopes \citep{Valotto97,Goto02,Yagi02,DePropris03}. Some studies seem to indicate, in fact,  that  the faint end slope of the LF is different in clusters and field, with the cluster environment being richer in faint galaxies than the field \citep{Popesso06,blanton05}.
However more recently \citet{agulli14} studying the spectroscopic LF of Abell 85 find that the faint-end slope of the LF is consistent with that of the field.

Finally, some studies claim that the cluster LF shows little variation across a wide range of cluster properties \citep{Colless89,Rauzy98,DePropris03,Popesso06}, while others find it to depend on cluster richness, Bautz-Morgan type \citep{Bautz70}, or distance from the cluster center \citep{Dressler78,Garilli99,Lopez97,Hansen05,Barkhouse07}. 

Differences in the estimated parameters might be related to the contamination from background galaxies, especially in the faint part of the LF, while the bright part can suffer from superposition of other clusters along the line of sight. While this second effect is more easily taken into account, since two sequences tend to appear in the cluster  Color--Magnitude diagram, the first one can be alleviated only by using statistical approaches, which have large uncertainties.

In this respect the availability of large galaxy surveys in the recent years has prompted the study of the global characteristics of galaxy clusters. In particular we have at our disposal the large sample of  the WIde-field Nearby Galaxy-clusters survey (hereafter WINGS,  \citealt{Fasano2006}) of low redshift clusters that is particularly suited to this purpose, since galaxies have been observed over a large field around the cluster center and with the needed accuracy (in the WINGS survey we have a FWHM average seeing of $\sim$1.2 arcsec that converts into a spatial resolution of 1.2-1.4 kpc for our range of redshift). 
A reliable object classification, as well as an excellent morphological completeness and a good spectroscopic coverage make this survey the ideal place where to study, in particular, cluster galaxies luminosity functions.

The purpose of this paper is to present the LF of 72 clusters of galaxies belonging to the WINGS survey, for which we possess reliable star/galaxy classification and magnitudes up to $V\sim22$. This will help us to understand whether the LF varies or not  as a function of cluster characteristics such as the X--ray luminosity or the velocity dispersion. In Section \ref{sec:data} we describe the data sample we used, in Section \ref{sec:lum} we derive the LF. In Section \ref{sec:prop} we present our conclusions about the universality of the fitted luminosity function and the dwarf galaxy population. Finally in Section \ref{sec:composite_lf} we present our LF for morphologically selected samples of galaxies and for a much more limited sample of spectroscopically confirmed cluster members.

Throughout this work we have used the cosmological parameters $H_0$ = 70 km $s^{-1}Mpc^{-1}$, $\Omega_0$ = 0.3 and $\Omega_{\Lambda}$ = 0.7.

\section{Data sample}\label{sec:data}
WINGS has been designed to derive the properties of galaxies in the cluster environment in the local Universe, and it is therefore of particular relevance in the context of studying the local LF. 
Here we briefly summarize the main survey characteristics. WINGS is a multi-wavelength project based on the analysis of deep wide--field images of nearby clusters selected from the X--ray flux--limited samples described in \citet{Ebeling1996,Ebeling1998,Ebeling2000}. Their location in the sky has been chosen to minimize the contamination from the  Galactic extinction ($|b|\ge20^{\circ}$). Cluster redshifts are in the range $0.04-0.07$.
All the available data for the WINGS survey are described in \citet{moretti14}, in particular the spectroscopic follow--up for 48 clusters \citep{Cava2009}, as well as the photometric data in the optical \citep{Varela09}, near infrared \citep{Valentinuzzi2009} and U band \citep{Omizzolo2013}. Stellar masses and star formation histories have been derived for the subsample of galaxies with spectroscopy \citep{Fritz2007,Fritz2010,Fritz2014}. 

One of the primary goals of the WINGS survey has been, since the beginning of the observations, the spectroscopical coverage of large areas in each of the sampled cluster.
\citet{Cava2009} illustrates the final WINGS spectroscopic sample, which is made of 6137 galaxies (in 48 clusters) observed with two telescopes (WHT for the north sample and AAT for the south sample) with a medium resolution setup ($6-9\AA$). The wavelength coverage ranges from $\sim 3800$ to $6800$ \AA.
For these galaxies we could determine redshifts (with a median error of $\sim 30$ km s$^{-1}$ ) and membership as described in the original paper.
In order to maximize the probability to observe galaxies at the cluster redshift, without biasing the cluster sample, targets were selected on the basis of their properties so that background galaxies (redder than the cluster red sequence) could be reasonably avoided. In particular the spectroscopic sample is made of galaxies with $V \leq 20$ (total magnitude), $V_{fiber} < 21.5$ and $(B-V)_{5 kpc} \leq 1.4$. This last cut has been then slightly varied from cluster to cluster in order to optimize the observational setup.
We then, {\it a posteriori},  calculated the spectroscopical completeness as the ratio of the number of spectra with a redshift determination with respect to the number of galaxies in the photometric catalog obeying to the previous criteria. This completeness is essentially independent of the distance from the cluster center (for most clusters) and of the magnitude \citep[see][for a complete analysis]{Cava2009}.

\subsection{Computing the luminosity function}\label{sec:comp_LF}
We used the Sextractor photometric catalog of WINGS galaxies described in \citet{Varela09} which refers to optical (B, V) photometry of 76 cluster of galaxies, either observed with the INT telescope at La Palma, or with the 2.2m ESO telescope at La Silla. For each detection we possess a star/galaxy classification based on the Sextractor stellarity index \citep{Sex} that leads to a sample of 394280 galaxies, 180952 unknown objects and 183792 stars. 

As described in \citet{Varela09}, this classification has been severely tested against other parameters and visually inspected, when possible. A careful analysis of the results demonstrates that the classification of galaxies is reliable up to $V\sim22$, while for fainter objects (up to $V\sim24$) no conclusion about the star/galaxy classification can be safely drawn. 
In particular, simulations show that a certain fraction of unclassified objects (variable with magnitude) had to be considered as made of galaxies (see Fig. 8 of \citealt{Varela09}). We took into account this effect, by adding to the number of detections classified as galaxies a fraction of unknown/galaxy objects calculated interpolating the \citealt{Varela09} points in Fig. 8, above V=21.5.
From now on in the paper the population referred to as galaxies is already corrected for this factor.
  
The characteristics of the galaxy population have been shown to vary with cluster--centric distance \citep{Christlein03,Hansen05,Popesso06}, this fact likely producing a bias when analyzing different dynamical regions of clusters.
 To overcome this problem and make meaningful comparisons between clusters with different size and richness, as previously done by \citet{Popesso06} and \citet{Barkhouse07}, we selected only galaxies located inside $0.5 \times r_{200}$, defined as the radius of a sphere with interior mean density 200 times the critical density of the Universe at that redshift. The quantity $r_{200}$ in Mpc has been calculated from the velocity dispersion and redshift $z$, taken from \citet{Cava2009} using the following equation \citep{Finn+2005,Poggianti+2006}:
\begin{equation}
r_{200}=1.73\times\frac{\sigma_v}{1000 km/s}\times\frac{1}{\sqrt{\Omega_\Lambda+\Omega_0(1+z)^3}}\times h^{-1}
\end{equation}
where $\sigma_v$ is the velocity dispersion of the cluster.
The velocity dispersion measurement was not available for 4 out of 76 clusters, and they have been therefore excluded from our analysis.

We used the V {\it AUTO}  magnitude from Sextractor and applied the k--correction using the recipe given in \citet{Poggianti1997}. The correction is calculated on the basis of the $(B-V)$ color of each galaxy (relative to an aperture of $\sim 10.8$ kpc), which is considered a proxy for the galaxy type.
We also took in account the photometric completeness as described in \citet{Varela09}. We used a fit to their Fig. 5 to derive the global completeness function for our clusters, and corrected each LF bin for this value. In what follows we then fitted only the magnitude range (different for each cluster) where the completeness was larger than $90 \%$. Table \ref{tab:fitpar} lists this limit for each cluster.
As for the field contribution, we used the number counts of extended sources in the ELaIS--S1 area, given in \citet{Berta06}.
 Before applying the statistical correction we scaled the number counts to the area covered by our observations, and in particular to the area where we estimated the LF (i.e. $0.5 \times r_{200}$).

The Brightest Cluster Galaxies (BCG) always form a distinct class of objects \citep{Fasano2010}, and therefore have been excluded from our sample of galaxies.

Errors on the calculated number density have been derived following \citet{Lugger1986} and \citet{Barkhouse07} as
\begin{equation}
\sigma_N=\frac{\sqrt{N_{nc}+N_f+1.69\cdot N_f^2}}{A}
\end{equation}
where N is the corrected number of galaxies in the given bin, after the completeness and field subtraction, $N_{nc}$ is the original number of galaxies, $N_f$ is the number count of the field galaxies in the given bin and A is the area in Mpc$^2$.
   \begin{figure*}
   \centering 
   \includegraphics[width=0.45\textwidth]{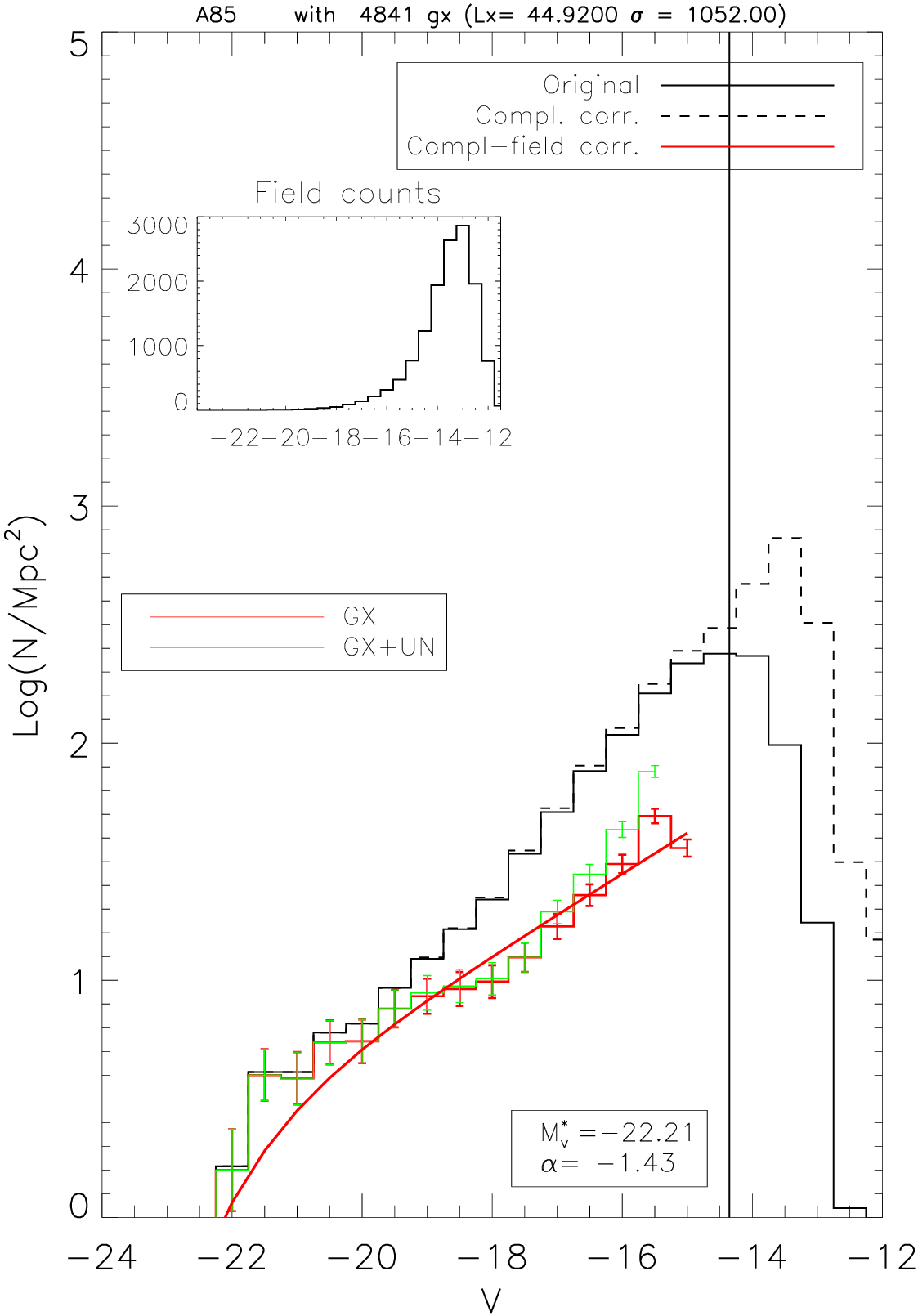}
   \includegraphics[width=0.45\textwidth]{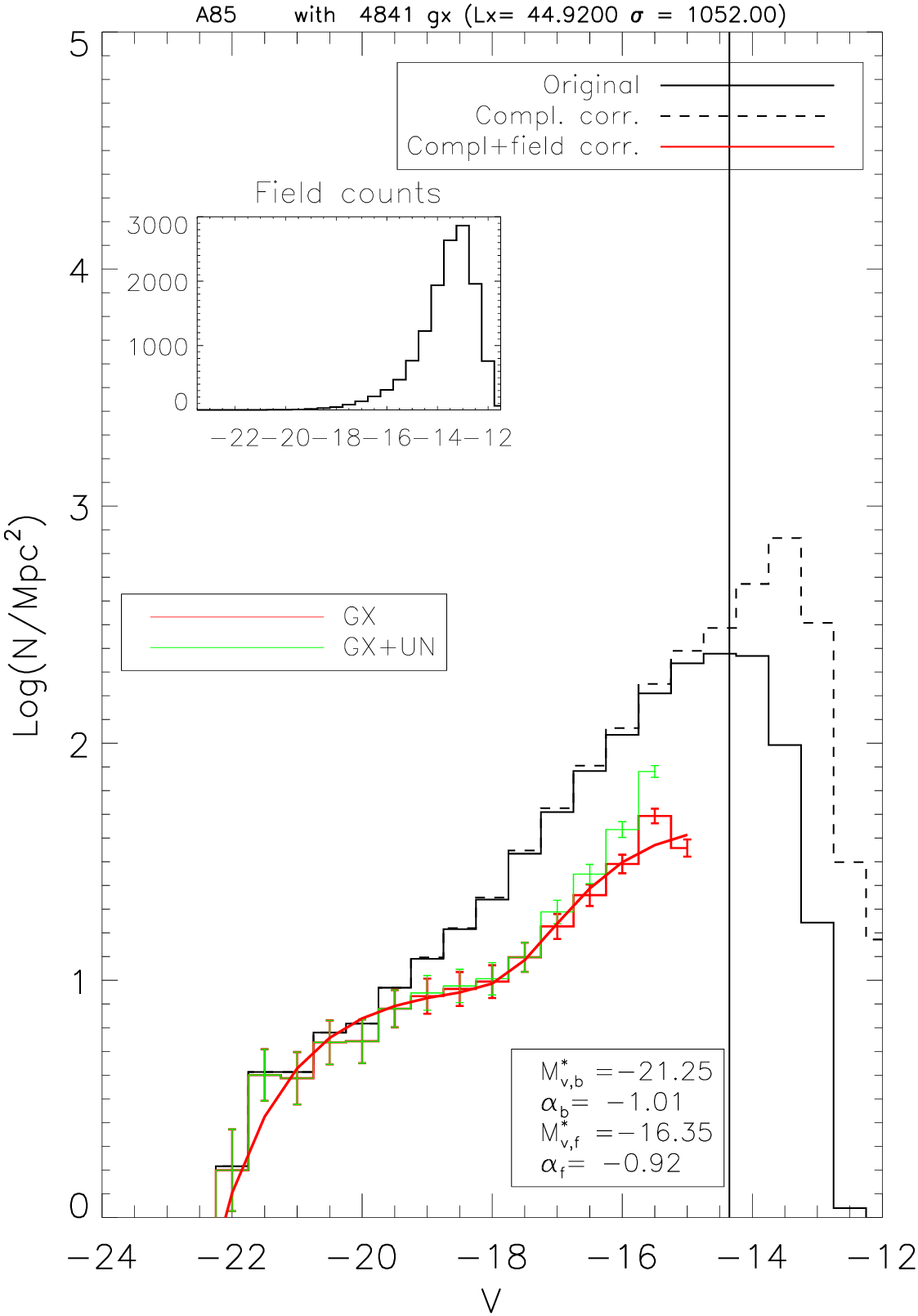}
      \caption{Luminosity function for the cluster A85: the black continuous line is the original LF, the dashed one is the same LF corrected for completeness. The vertical line shows the magnitude limit (different for each cluster) at which the completeness is $90 \%$. In red and green we show the LF of the two subsamples of galaxies and galaxies and unknown object, respectively. These two last distributions have been corrected for field contamination (whose number counts are shown in the inset). Superimposed to the red LF is the best fit that we obtained using a single Schechter function (left panel) and a double Schechter function (right panel). In the bottom right insets we give the relative parameters.}
         \label{fig:lf1s_a85}
   \end{figure*}

We also derive luminosity functions for galaxies having different morphological classes (Ellipticals, S0 and later types).
For a subsample of 39124 galaxies we were able to perform an automatic morphological classification using MORPHOT \citep{Fasano+2012}, a tool that has been created for the WINGS survey. The classification is based on 21 visual diagnostics and on a parallel Neural Network machine. We refer the reader to the original paper for details on the tool.
The MORPHOT ability to classify objects obviously depends on the cluster distance, as well as on the overall photometric quality of observations. We therefore decided to use only galaxies having magnitudes brighter than the one where the MORPHOT completeness is higher than 0.5 (the MORPHOT completeness is defined as the ratio between the number of galaxies classified and the number of photometric detections classified as galaxies).
This limit is obviously variable within the cluster sample but is in the interval $M_V=-16.5 - -17.5$. In particular $\sim$18\% of the cluster sample has a limit of completeness of 50\% at magnitude $M_V=-16.0$, for the 40\% the same limit is at $M_V=-17.0$ and the remaining of the sample  reach the 50\% completeness at $M_V=-17.5$.  The number counts of galaxies in each morphological class have been corrected for the morphological incompleteness.

For the LF of different morphological classes, we decided to use the sample described in \citet{Calvi+2011b}, derived from the Millennium Galaxy Catalogue (MGC) by \citet{Liske+03,Driver+05} to perform a meaningful background subtraction. The sample is made of 3210 galaxies located in the so called "general field" (see \citealt{Calvi+2011a} for details about the subsample definitions) for which the morphological classification has been performed using MORPHOT.

We first calculated the morphological mix of galaxies in each magnitude bin, and then rescaled this number to the total number of galaxies expected in that bin from the number counts by \citet{Berta06}.

Finally we construct the spectroscopic LF for the subsample of 21 clusters with a spectroscopic completeness higher than 50\%. For this we used the spectroscopic information given in \citet{Cava2009} to derive the membership of our detections, and corrected for spectroscopic incompleteness as described in \citet{Cava2009} and \citet{Vulcani+2011b}.

\section{Cluster Luminosity Functions}\label{sec:lum}
\subsection{Single Schechter function fit}\label{sec:single_lf}
For each cluster we calculated the LF as described in the previous section for three different classes of objects (galaxies, stars and unknown). As an example in Fig.\ref{fig:lf1s_a85} (for all the clusters in the on line version of the paper) we show the results for the cluster A85.  The LF for galaxies is represented by the red line histogram, while that  for galaxies plus all unknown is represented by the green line histogram.

Each LF has then be fitted up to the limiting magnitude (vertical line in Fig.\ref{fig:lf1s_a85}), defined as the magnitude at which the sample is 90\% complete. This number varies with the cluster distance and the quality of observations.
In Figure \ref{fig:lf1s_a85} it can be seen how the completeness correction and the field subtraction act on the final LF (dashed black line and red/green lines, respectively).
The left panel of Fig.\ref{fig:lf1s_a85} shows the best fit of the galaxy LF obtained using one single Schechter \citep{Schechter1976} function of the form:

\begin{equation}
\phi(L)=\phi^\ast \left[\left(\frac{L}{L^\ast}\right)^{\alpha} exp \left(\frac{-L}{L^\ast}\right)\right]
\end{equation}
that describes the number of galaxies per unit volume ($\phi$) as a function of the galaxy luminosity L, the characteristic galaxy luminosity $L^\ast$, corresponding to the knee of the LF, and the slope of the LF at low luminosities $\alpha$.
If we let free to vary the Schechter parameters, we obtain unphysical results in clusters that have a poor galaxy population, or where a hint for the presence of a secondary sequence of a background cluster is present.
We excluded from the calculation of the mean/median Schechter parameters fit these clusters (49/72), i.e. those having errors in the derived $M_V^\ast$ and $\alpha$ larger than 2.0 and 0.275, respectively.

\begin{figure}
   \centering
   \includegraphics[width=0.45\textwidth]{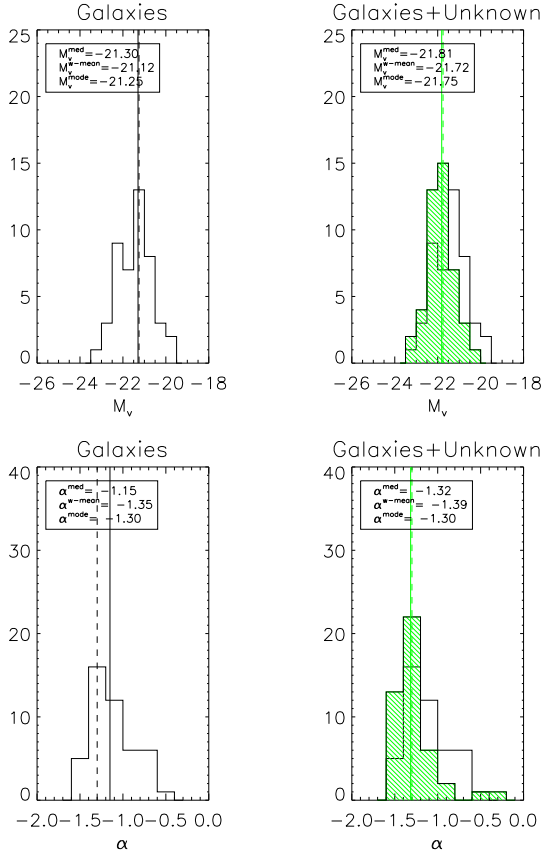}
      \caption{Distribution of $M_V^\ast$ (upper panels) and $\alpha$ (lower panels) derived by fitting one single Schechter function to our LF. In all plots the black histogram shows the results we obtained by analyzing only objects classified as galaxies while green histogram to the population of galaxies and unknown objects.}
         \label{fig:lf1sall}
   \end{figure}

Fig. \ref{fig:lf1sall} shows the distributions of $M_V^\ast$ (upper panels) and $\alpha$ parameters (lower panels) for two subsamples of objects, i.e. galaxies (black continuous line) and galaxies plus unknown (superimposed as green dashed histogram on the right panel).
   Our median (mode) values for the luminosity function characteristic luminosity and slope are $M_V^\ast=-21.30$ $(-21.25)$ and $\alpha=-1.15$ $(-1.30)$  considering the sample of galaxies, whereas $M_V^\ast$ becomes brighter ($-21.81,-21.75$ for median and mode)  including also unknown objects. 
We also calculated the weighted mean of $M_V^\ast$ and $\alpha$ using as weight the error on the derived quantity, obtaining $M_V^\ast=-21.12 (-21.72)$ and $\alpha=-1.35 (-1.39)$ for the two subsamples of pure galaxy population and galaxy plus unknown. In this case we did not exclude clusters with a poor determination of the parameters.

\begin{figure}
   \centering
   \includegraphics[width=0.5\textwidth]{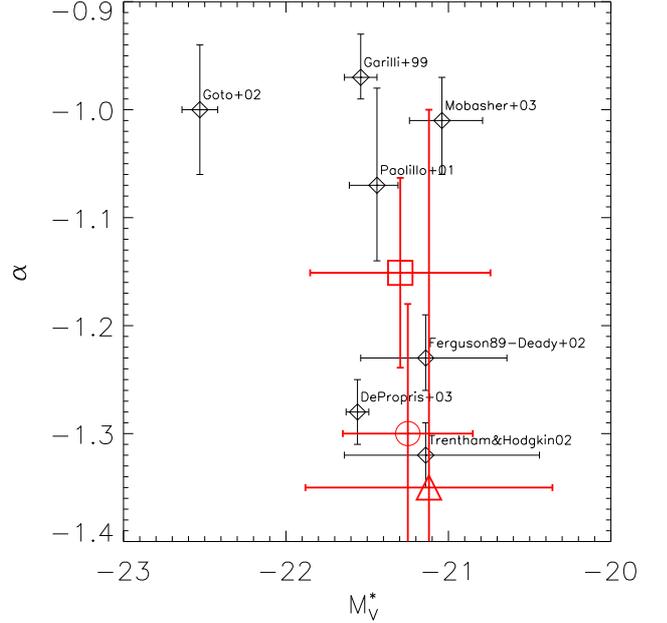}
      \caption{Comparison with literature data, homogenized to the same photometric band and cosmological parameters. The squared point refers to the WINGS median values, the circle is the mode, while the triangle is the weighted mean. Errors are the mean errors in the derived parameters.}
         \label{fig:lf1_cfrlit}
   \end{figure}

To compare these results with literature data we considered first the Virgo, Fornax and the 2dFGRS surveys \citep{trentham2002,ferguson1989,deady+2002,DePropris03} of nearby clusters. To make this comparison we transformed their B band data to our V band using a value of $(B-V)=1$ (that is the typical color of a Single Stellar Population with an age larger than $\sim 6$ Gyr, with solar metallicity and Salpeter IMF, see \citealt{bc03} models) and took into account the different cosmology. We find that our estimates are in agreement with the literature where the LF has been calculated up to a very faint magnitude limit, as shown in Fig.\ref{fig:lf1_cfrlit}. 
There are, however, different results in the literature, e. g. the ones coming from Coma \citep{mobasher+2003} and other clusters \citep{Garilli99,Goto02,Paolillo+2001}, where the slope turns out to be shallower than the one found in Virgo, Coma and the 2dFGRS survey. In fact, after having converted the data from \citet{Garilli99,Goto02,Paolillo+2001} using the relation $B=g+0.54$ \citep{Liske+03}, and the (B-V) color term described above, we find a value for $M_V^\ast$ in broad agreement with all the data but \citealt{Goto02}.

The slope is in agreement with studies based on fields of similar size \citep{trentham2002,ferguson1989,deady+2002,DePropris03}, while it turns out to be steeper than the one found for core regions \citep{Garilli99,Goto02,Paolillo+2001}, where evolutionary processes build up the cD galaxy leading to the disruption of dwarf galaxies. Coma \citep{mobasher+2003} lies in the region of shallower slopes, but this can be due to selection effects, since the spectroscopic sample is based on the R-magnitude, while the LF refers to the B-band \citep{driver+depropris03}.

Our magnitude limit lies between $M_V=-13.6$ and $M_V=-15.15$, which is the limiting magnitude for the Coma data here considered. Virgo and Fornax have even deeper magnitude limits, and are fitted with nearly the same $M_V$.  \citet{Paolillo+2001} sample has a limiting magnitude of $M_V=-17$, and \citet{Garilli99,DePropris03} sample reaches $M_V=-18$. There is, therefore, the possibility that in these last clusters the rising faint end of the LF is not visible, thus making their Schechter's slope $\alpha$ flatter.
The mean errors in the derived fit are $0.55$ and $0.09$ in $M_V$ and $\alpha$, respectively.

\subsection{Double Schechter function fit}\label{sec:lf2s}
The left panel of Fig. \ref{fig:lf1s_a85}  clearly shows that a single Schechter fit does not reproduce the details of the LF, in particular the steepening of the faint end of the LF and the central plateau.

Recent studies on nearby clusters \citep[see][among others]{Boselli+08,Penny+2011,agulli14} have indeed confirmed that the LF steepens at faint magnitudes, especially when moving towards the external regions of the cluster.

Therefore, we  fit our LFs using a  double Schechter function \citep[see][among others]{Driver+94,Hilker+03,gonzalez+06,Popesso06, Barkhouse07}. The function has the following form:

\begin{equation}
\phi(L)=\phi^\ast \left[\left(\frac{L}{L_b^\ast}\right)^{\alpha_b} exp \left(\frac{-L}{L_b^\ast}\right)+\left(\frac{L_b^\ast}{L_f^\ast}\right)\times\left(\frac{L}{L_f^\ast}\right)^{\alpha_f} exp\left(\frac{-L}{L_f^\ast}\right)\right]
\end{equation}
where the number of galaxies per unit volume $\phi$ depends both on the characteristic magnitude and slope in the bright part of the LF ($L_b^\ast$ and $\alpha_b$, respectively) and on the characteristic magnitude and slope in the faint part ($L_f^\ast$ and $\alpha_f$, respectively).
Table \ref{tab:fitpar} lists the results of our fits for all clusters. In column 1 we identify the cluster, in cols. 2 and 3 we give the area (in Mpc) over which the luminosity function has been calculated and the dwarf to giant ratio (DGR) respectively. The last quantity has been calculated as the ratio between the number of objects with absolute V magnitude brighter than -19.0 and the number of object fainter than this limit (see \citealt{poggianti+2001}) but brighter than -15.15, which is the faintest magnitude limit reached in all clusters.
In columns 4 to 11 we give the parameters of the best fitting double Schechter fit  $M_V^b$, Err($M_V^b$), $\alpha^b$, Err($\alpha^b$), $M_V^f$, Err($M_V^f$), $\alpha^f$ and Err($\alpha^f$).
Finally, the last four columns give the total number of galaxies analyzed $N_{gx}$, the cluster velocity dispersion $\sigma_v$ (in $km s^{-1}$), the cluster X--ray luminosity $L_X$ and the absolute magnitude limit $M_{lim}$, up to which the LF has been fitted.

The last six rows give the median and mode results for two subsamples: the one including only objects classified as galaxies (corrected for the fraction of unknown that can be classified as galaxies), and the one where all objects (i.e. galaxies and unknown objects, excluding stars) are included. For both subsamples we give the parameters of the free fitting, and the parameters of the fit obtained by imposing $\alpha_b=-1.10$.

We first fitted the double Schechter function to each LF letting all parameters free, and then considered good fits those with errors in the magnitudes lower than $2.5$ and errors in the slopes lower than $1.0$. We were able to fit 41/72 clusters and obtained median values of $-21.15$ and $-16.30$ for the bright and faint end $M_V^{\ast}$, while for the slopes the values are $-0.97$ and $-0.6$, respectively. These values together with the mode values are given in Tab.\ref{tab:fitpar}.

We compared our results with the ones by \citealt{Popesso06} and \citealt{Barkhouse07}, after having transformed their magnitude values to the V band (and using our cosmology). For the values given in \citet{Popesso06} we converted the {\it g} magnitude using the transformation $V=g-0.565(g-r)$, while for the ones taken from \citet{Barkhouse07} we used $(B-V)=1$ and $(B-R)=1.8$. This last value is the mean color calculated by \citet{lopez+04} for the same clusters analyzed in \citet{Barkhouse07} at $R=17$.
Both \citet{Popesso06} and \citet{Barkhouse07} calculated their LFs inside the same physical region in each cluster (i.e. $r_{200}$ or $r_{500}$ in the first case, and between $0.2$ and $0.4$ $r_{200}$ and $0.4$ and $0.6$ $r_{200}$ in the second case).

\begin{figure}
   \centering
   \includegraphics[width=0.5\textwidth]{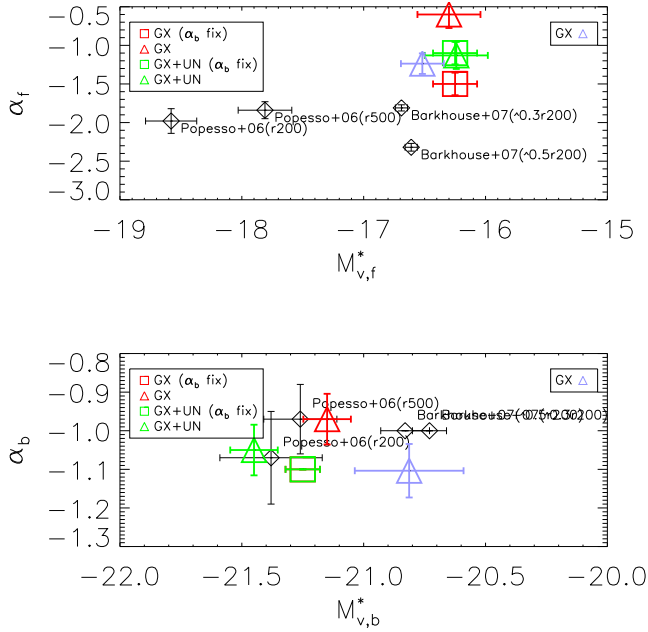}
      \caption{Comparison with literature data, homogenized to the same photometric band and cosmological parameters. Squared point refer to the WINGS median values of the galaxies  (including the unknown galaxies) subsample, the triangle to the subsample all objects of galaxies and unknown sources, excluding stars. When fixing the $\alpha_b$ the $M_V^b$ is coincident in the two cases considered (i.e. population of pure galaxies or galaxies and unknown objects), therefore only one (green) square remains visible. The purple triangle refers to the weighted mean of the population of pure galaxies, derived leaving free to vary the bright end slope of the LF.}
         \label{fig:lf2_cfrlit}
   \end{figure}

When compared with these data, our results seem to favor a brighter ($0.35$-$1.0$ mag) magnitude for the bright characteristic magnitude of the LF and a fainter one for the faint end part ($0.4$-$0.9$ mag, see fig.\ref{fig:lf2_cfrlit}). At the same time the slope in the bright regime is compatible with the values given in literature, while it is flatter in the faint end regime (see, for example, values given in \citealt{Popesso06,Barkhouse07} but also \citealt{boue+2008} for results more similar to ours).

In order to better compare our findings with others, we run a fit after having imposed a bright end slope of $\alpha_b=-1.10$, corresponding to the mode value of our LFs fits and to the value found by \citet{Popesso06} within $r_{200}$. In this way we were able to fit 56/72 clusters (selected using the same criterion described above).

Fig.\ref{fig:lf2_cfrlit} shows the comparison between our derived parameters (median values) for the LF faint end (upper panel) and for the bright end (lower panel).
In both cases we show in red the median values for the galaxy (including the unknown galaxies) population, and in green the population of galaxies and unknown object, without the stellar component.
Squares refer to values derived using a fixed $\alpha_b=-1.10$ and triangles to values derived leaving free all parameters.
The purple triangle shows, finally, the weighted mean of the fitted parameters, that we calculated only for the subsample of galaxies and leaving free the bright end of the LF. If considering the weighted mean, the bright part of the LF shows a much better agreement with the values derived by \citealt{Barkhouse07}, but in the faint part the results show again a flatter slope.

  \begin{figure*}
   \centering
   \includegraphics[width=0.45\textwidth]{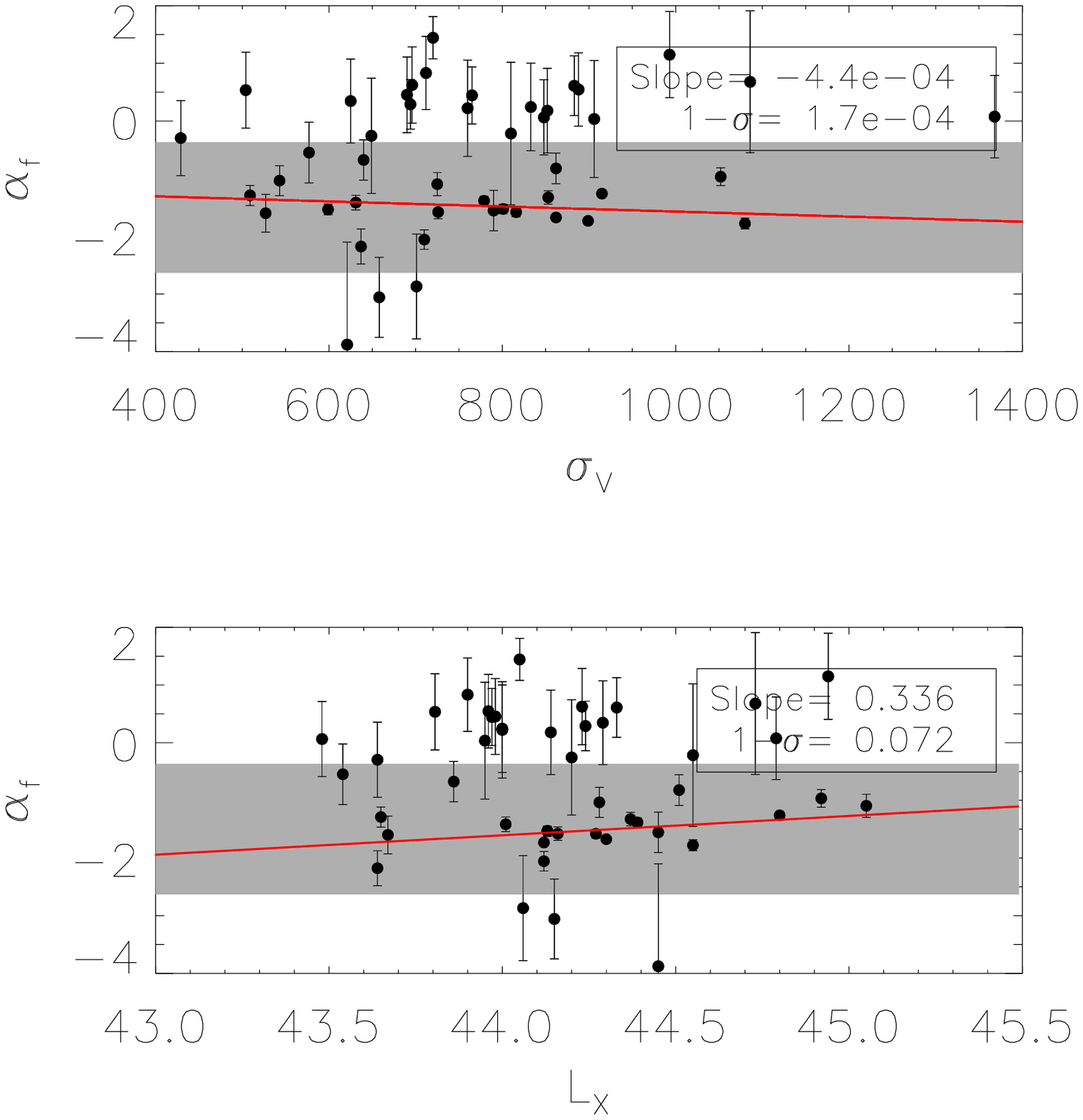}
   \includegraphics[width=0.45\textwidth]{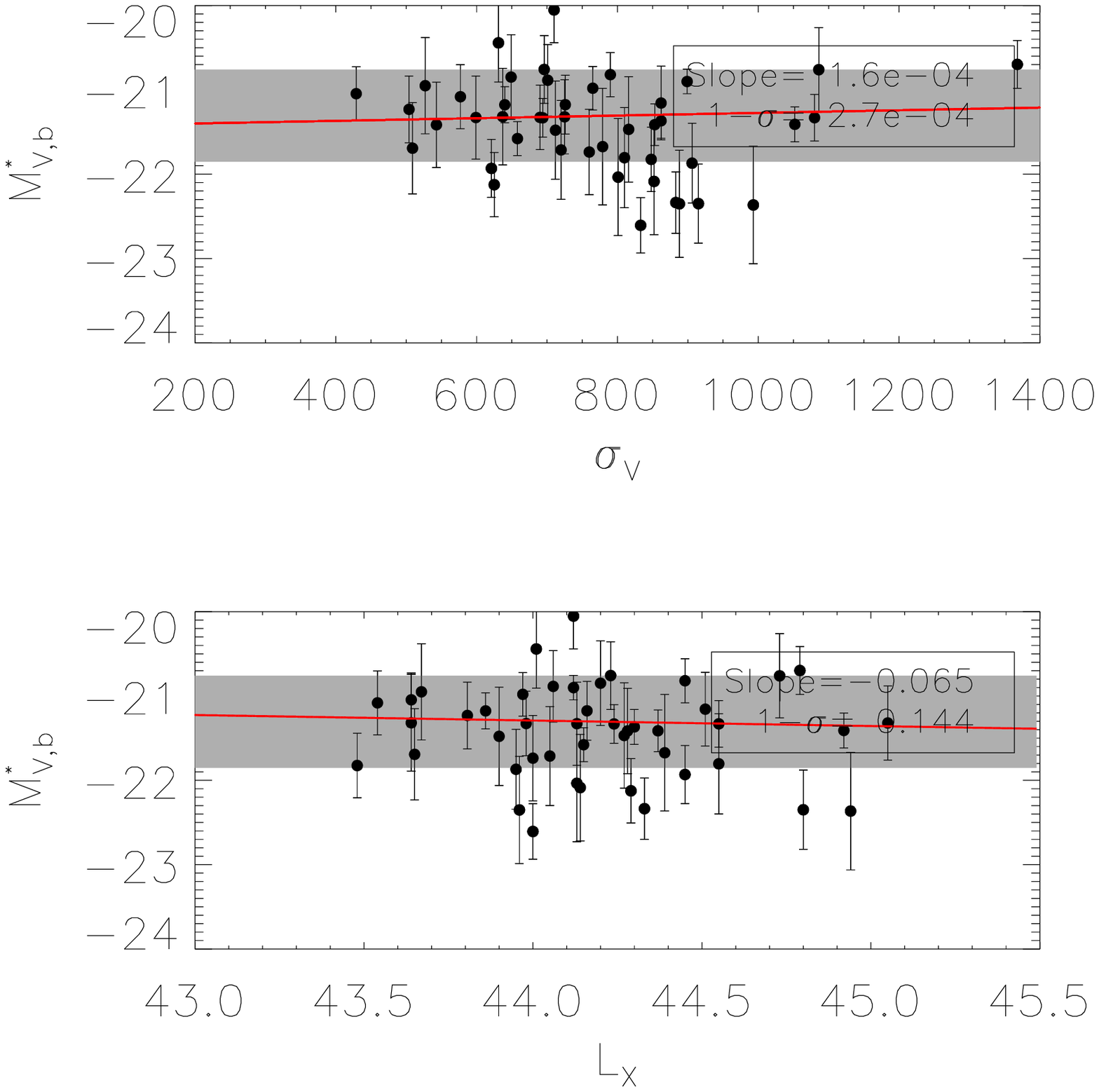}
      \caption{Variation of $\alpha_f$ (left panels) and $M_V$(right panels) with $\sigma_v$ (upper panel) and $L_X$ (lower panel), with superimposed the least square fit.}
         \label{fig:lf_pars}
   \end{figure*}
 The derived best fit parameters show a good agreement in the bright part of the LF, where both samples of pure galaxy population and the one including unknown objects have $M^{\ast}_{V,b}=-21.25$ and a slope of -1.10. 
As for the faint part of the LF, we find fainter characteristic magnitudes and slightly flatter slopes, even after having fixed the bright end slope.

The net effect of including in the WINGS sample more galaxies taken from the unknown class is to have a brighter characteristic magnitude in the bright end part of the LF and more or less the same slope in the faint end part.
This result is somewhat unexpected, since unknown galaxies included in the second sample are mainly dwarf galaxies, that should have the effect, if any, to steepen the faint end LF. 
However, the double Schechter fit tries to fit simultaneously the two parts, so that in order to better reproduce the steepening of the faint end it also moves the bright end magnitude towards the faint end. What we find, in fact, are LF flatter than those found so far, but with a more pronounced central plateau.

The main concern in the Schechter fitting is related to the large errors on the single fits, that have been derived leaving free to vary all parameter of the double Schechter function (or fixing one of them, the bright end slope), as can be seen from Table \ref{tab:fitpar}. This statistical effect is known, and can be solved by constructing a composite LF, where all clusters contribute, thus giving much stronger constraints on the resulting LF in particular for their faint end.
However, the composite LF is meaningful only in the case in which we think that the cluster LF is universal, otherwise differences would be canceled out and the derived parameters would be a sort of average behavior. Next section is dedicated to a more detailed discussion on the universality of the WINGS cluster LF.

 \begin{table*}[htdp]
\scriptsize
\caption{Schechter Function Parameters: for each cluster (col.1) we give the area covered by our observations, in Mpc$^2$ (col. 2), the Dwarf--to--Giant ratio (col. 3), the fitted $M_V^b$ and $\alpha^b$ (col. 4, 6) and the relative error on the fit (col. 5, 7), the fitted $M_V^f$ and $\alpha^f$ (col. 8, 10) and the relative error on the fit (col. 9, 11), the $\chi^2$ of the fit (per degree of freedom),  the number of galaxies, the cluster velocity dispersion, the (log of) X-ray luminosity in the range 0.1-2.4 keV from \citet{Ebeling1996} and the limiting absolute magnitude.}
\begin{center}
\begin{tabular}{|c|c|c||c|c|c|c||c|c|c|c||c|c|c|c|c|}
\hline
Cluster & Area & DGR &  $M_V^b$ & Err($M_V^b$) & $\alpha^b$ & Err($\alpha^b$) &$M_V^f$ & Err($M_V^f$) & $\alpha^f$ & Err($\alpha^f$)& $\chi^2$& $N_{gx}$ & $\sigma_v$ & $L_X$ & $M_{V,lim}$\\
\hline
A85      &   3.654 & 4.10  &    -21.25 &    0.30 &   -1.01 &    0.13 &  -16.35 &    0.28 &   -0.92 &    0.16 &     0.42 & 4841 & 1052 &  44.92  &  -14.35\\
A119     &   2.454 & 4.08  &    -20.13 &    0.44 &   -0.61 &    0.25 &  -15.82 &    0.34 &   -0.60 &    0.21 &    2.25 & 5305 & 862 &   44.51  &  -14.00\\
A133     &   2.730 & 2.23  &    -21.42 &    1.22 &   -0.92 &    0.72 &  -17.31 &    1.36 &   -0.40 &    1.09 &     1.17 & 2599 & 810 &   44.55  &  -14.62\\
A147     &   1.640 & 2.80  &    -32.37 &    99.99&    0.45 &    99.99&  -22.20 &    8.74 &   -1.08 &    0.18 &    7.61 & 3181 & 666 &   43.73  &  -13.96\\
A151     &   2.382 & 3.14  &    -21.59 &    2.68 &   -0.87 &    2.26 &  -19.32 &    1.01 &   -0.99 &    0.30 &    3.00 & 3480 & 760 &   44.00  &  -14.35\\
A160     &   1.362 & 2.17  &    -22.39 &    2.15 &   -0.96 &    0.39 &  -15.01 &    99.99&  -43.67 &    99.99&    5.90 & 2381 & 561 &   43.58  &  -14.13\\
A168     &   1.045 & 3.26  &    -22.57 &    1.61 &   -1.06 &    0.19 &  -16.30 &    0.48 &   -0.97 &    0.26 &    1.51 & 2367 & 503 &   44.04  &  -14.00\\
A193     &   2.258 & -0.80 &    -20.10 &    0.54 &   -0.16 &    0.47 &  -15.82 &    99.99&   -2.44 &   99.99 &    1.39 & 3215 & 759 &   44.19  &  -14.22\\
A376     &   2.371 & 2.96  &    -21.70 &    1.14 &   -1.02 &    0.35 &  -15.81 &    1.08 &    0.04 &    1.29 &    1.73 & 3633 & 852 &   44.14  &  -14.24\\
A500     &   1.835 & 2.32  &    -21.77 &    0.34 &   -1.15 &    0.06 &  -15.23 &    0.09 &   -3.49 &    1.12 &     0.41 & 2157 & 658 &   44.15  &  -14.98\\
A548b    &   2.672 & 2.43  &    -21.37 &    0.56 &   -0.92 &    0.34 &  -16.55 &    0.87 &   -0.24 &    0.72 &    1.33 & 6037 & 848 &   43.48  &  -13.93\\
A602     &   1.982 & 5.14  &    -21.15 &    0.87 &   -0.92 &    0.33 &  -15.84 &    0.31 &    1.05 &    0.68 &    2.05 & 1845 & 720 &   44.05  &  -14.81\\
A671     &   3.143 & 1.30  &    -20.68 &    0.55 &   -0.02 &    0.64 &  -18.62 &    0.61 &   -0.33 &    0.39 &     0.90 & 3237 & 906 &   43.95  &  -14.33\\
A754     &   3.580 & 1.63  &    -19.82 &    0.33 &    0.76 &    0.41 &  -17.46 &    0.32 &    0.43 &    0.32 &    1.61 & 3315 & 1000 &  44.90  &  -14.55\\
A780     &   0.975 & 0.50  &    -20.84 &    1.04 &   -0.24 &    0.76 &  -18.52 &    0.37 &   -0.08 &    0.25 &     0.73 & 502 & 734 &    44.82  &  -14.55\\
A957x    &   2.050 & 1.73  &    -18.91 &    0.27 &    2.00 &    0.55 &  -16.52 &    0.33 &    1.44 &    0.63 &    1.78 & 2024 & 710 &   43.89  &  -14.04\\
A970     &   2.350 & -0.30 &    -21.21 &    0.43 &   -0.85 &    0.14 &  -11.32 &    0.00 &   -2.49 &    0.00 &     0.98 & 1227 & 764 &   44.18  &  -14.69\\
A1069    &   1.955 & 3.20  &    -21.69 &    0.91 &   -1.19 &    0.19 &  -16.23 &    0.46 &    0.72 &    0.88 &     0.95 & 1522 & 690 &   43.98  &  -14.88\\
A1291    &   0.794 & 4.30  &    -21.01 &    0.91 &   -1.09 &    0.43 &  -16.39 &    1.39 &   -0.32 &    1.42 &    1.18 & 1045 & 429 &   43.64  &  -14.23\\
A1631a   &   1.715 & 3.08  &    -21.74 &    0.37 &   -1.22 &    0.07 &  -15.00 &    0.45 &   -0.81 &    0.43 &     0.57 & 3987 & 640 &   43.86  &  -14.13\\
A1644    &   3.120 & 3.77  &    -21.96 &    0.52 &   -1.29 &    0.10 &  -16.00 &    0.23 &   -2.12 &    0.26 &    0.89 & 9128 & 1080 &  44.55  &  -14.22\\
A1668    &   1.751 & 1.10  &    -20.30 &    0.36 &   -0.87 &    0.13 &  -11.40 &    0.00 &   -2.42 &    0.00 &    1.21 & 1455 & 649 &   44.20  &  -14.78\\
A1736    &   2.401 & 3.32  &    -22.25 &    0.41 &   -1.30 &    0.07 &  -15.56 &    0.29 &   -1.72 &    0.30 &    0.69 & 7680 & 853 &   44.37  &  -14.14\\
A1795    &   2.191 & 3.90  &    -21.72 &    0.59 &   -1.37 &    0.07 &  -15.01 &    99.99&  -42.05 &   99.99 &     0.86 & 3662 & 725 &   45.05  &  -14.71\\
A1831    &   1.221 & 4.14  &    -23.08 &    1.43 &   -1.45 &    0.08 &  -16.42 &    1.16 &    0.77 &    2.02 &    2.82 & 1730 & 543 &   44.28  &  -14.73\\
A1983    &   1.173 & 2.97  &    -22.99 &    2.48 &   -1.36 &    0.10 &  -14.66 &    0.49 &   -3.16 &    2.10 &    1.68 & 3117 & 527 &   43.67  &  -13.97\\
A1991    &   1.500 & 5.27  &    -20.29 &    0.45 &   -0.50 &    0.35 &  -16.84 &    0.26 &   -1.22 &    0.13 &    1.03 & 2770 & 599 &   44.13  &  -14.59\\
A2107    &   1.477 & 1.03  &    -20.33 &    0.37 &   -0.68 &    0.19 &  -13.30 &   99.99 &   -2.35 &    99.99&    1.07 & 2650 & 592 &   44.04  &  -13.88\\
A2124    &   2.622 & 5.99  &    -22.03 &    1.27 &   -1.10 &    0.28 &  -17.17 &    0.43 &   -1.52 &    0.24 &    1.14 & 3362 & 801 &   44.13  &  -14.86\\
A2149    &   0.532 & 3.74  &    -19.63 &    0.72 &    0.17 &    0.87 &  -17.11 &    0.45 &    0.36 &    0.49 &    7.91 & 444 & 353 &    43.92  &  -14.84\\
A2169    &   1.151 & 3.16  &    -24.19 &    4.66 &   -1.45 &    0.08 &  -15.05 &    99.99&  -39.87 &   99.99 &     0.85 & 1599 & 509 &   43.65  &  -14.51\\
A2256    &   4.612 & 1.57  &    -21.74 &    0.32 &   -1.11 &    0.07 &  -12.22 &    0.00 &   -1.93 &    0.00 &     0.74 & 4264 & 1273 &  44.85  &  -14.63\\
A2271    &   1.066 & 1.78  &    -20.90 &    0.94 &   -0.96 &    0.48 &  -16.65 &    0.56 &    0.33 &    0.96 &     1.24 & 546 & 504 &    43.81  &  -14.62\\
A2382    &   3.250 & 1.42  &    -21.41 &    1.30 &   -0.69 &    1.12 &  -18.94 &    0.77 &   -0.40 &    0.72 &     1.03 & 3015 & 888 &   43.96  &  -14.89\\
A2399    &   2.119 & 1.07  &    -20.84 &    0.41 &   -0.76 &    0.16 &  -14.43 &    99.99&   -3.12 &   99.99 &    1.09 & 2066 & 712 &   44.00  &  -14.58\\
A2415    &   1.909 & 4.27  &    -20.56 &    0.79 &   -1.03 &    0.30 &  -14.80 &    0.51 &    0.51 &    0.87 &    1.99 & 1935 & 696 &   44.23  &  -14.67\\
A2457    &   1.492 & 0.31  &    -20.79 &    0.67 &    0.57 &    0.88 &  -19.63 &    1.04 &   -0.54 &    0.42 &    1.72 & 1011 & 580 &   44.16  &  -14.76\\
A2572a   &   1.565 & 4.91  &    -22.46 &    1.13 &   -1.41 &    0.10 &  -15.30 &    0.69 &   -2.02 &    0.66 &    1.03 & 4089 & 631 &   44.01  &  -13.81\\
A2589    &   2.141 & 5.28  &    -35.96 &   99.99 &   -1.58 &    0.06 &  -14.40 &   99.99 &  -24.93 &   99.99 &    0.85 & 7211 & 816 &   44.27  &  -13.83\\
A2593    &   1.765 & 1.74  &    -20.83 &    0.68 &   -1.10 &    0.14 &  -14.27 &    0.28 &   -1.64 &    1.09 &    2.80 & 4946 & 701 &   44.06  &  -13.87\\
A2622    &   2.075 & 2.34  &    -22.19 &    1.69 &   -1.18 &    0.32 &  -16.95 &    1.07 &    0.21 &    1.76 &     0.96 & 1893 & 696 &   44.03  &  -14.77\\
A2626    &   1.610 & 0.90  &    -21.54 &    1.03 &   -0.84 &    0.72 &  -17.59 &    0.87 &    0.07 &    1.11 &     0.77 & 1297 & 625 &   44.29  &  -14.55\\
A2657    &   0.634 & 1.64  &    -20.66 &    0.86 &   -0.44 &    0.39 &  -14.22 &    4.52 &   -0.46 &    9.17 &    5.37 & 916 & 381 &    44.20  &  -14.04\\
A2717    &   1.267 & 0.53  &    -21.30 &    1.93 &   -0.91 &    0.49 &  -15.51 &   99.99 &  -21.59 &   99.99 &    2.60 & 1815 & 553 &   44.00  &  -14.17\\
A2734    &   1.260 & 1.29  &    -20.59 &    0.67 &   -0.76 &    0.35 &  -14.81 &   11.15 &   -0.83 &   24.88 &    2.05 & 1488 & 555 &   44.41  &  -14.68\\
A3128    &   2.980 & 2.63  &    -21.91 &    0.71 &   -0.97 &    0.29 &  -16.41 &    0.57 &    0.36 &    0.86 &    1.24 & 3546 & 883 &   44.33  &  -14.61\\
A3158    &   4.544 & 2.14  &    -20.83 &    0.31 &   -0.99 &    0.08 &   -7.91 &    0.00 &   -3.19 &    0.00 &    1.02 & 5911 & 1086 &  44.73  &  -14.57\\
A3266    &   4.979 & 3.52  &    -20.41 &    0.55 &   -0.99 &    0.26 &  -15.13 &    0.83 &   -0.06 &    0.90 &    0.90 & 6737 & 1368 &  44.79  &  -13.58\\
A3376    &   1.593 & 7.01  &    -21.17 &    0.79 &   -0.78 &    0.40 &  -17.58 &    0.40 &   -1.29 &    0.12 &    1.74 & 4028 & 779 &   44.39  &  -14.12\\
A3395    &   2.684 & 3.37  &    -20.80 &    0.46 &   -1.09 &    0.16 &  -15.53 &    0.24 &   -1.54 &    0.53 &    1.26 & 4555 & 790 &   44.45  &  -14.52\\
A3490    &   2.000 & 2.19  &    -21.07 &    0.50 &   -0.94 &    0.35 &  -18.15 &    0.67 &   -0.22 &    0.85 &     0.68 & 1652 & 694 &   44.24  &  -15.15\\
A3497    &   2.099 & 5.67  &    -21.55 &    0.71 &   -1.23 &    0.22 &  -16.96 &    0.39 &   -1.70 &    0.29 &     0.91 & 3863 & 726 &   44.16  &  -15.07\\
A3528a   &   2.114 & 6.23  &    -21.29 &    0.27 &   -1.27 &    0.09 &  -16.71 &    0.14 &   -1.90 &    0.12 &     0.36 & 6533 & 899 &   44.12  &  -14.55\\
A3528b   &   2.053 & 6.18  &    -21.62 &    0.40 &   -1.22 &    0.12 &  -16.97 &    0.21 &   -1.78 &    0.13 &     0.71 & 4829 & 862 &   44.30  &  -14.53\\
A3530    &   1.318 & 2.54  &    -20.85 &    0.61 &   -0.92 &    0.24 &  -14.50 &    3.44 &   -0.38 &    5.97 &    2.39 & 1638 & 563 &   43.94  &  -14.60\\
A3532    &   1.611 & 2.39  &    -22.41 &    0.73 &   -1.19 &    0.08 &  -15.09 &    0.21 &   -8.22 &   19.37 &    1.08 & 2396 & 621 &   44.45  &  -14.65\\
A3556    &   1.305 & 2.89  &    -20.57 &    0.61 &   -0.49 &    0.30 &  -15.44 &    0.60 &   -0.51 &    0.67 &    3.15 & 2922 & 558 &   43.97  &  -14.24\\
A3558    &   2.815 & 2.85  &    -22.42 &    1.89 &   -1.21 &    1.90 &  -21.27 &    0.64 &   -1.22 &    1.78 &     0.53 & 6934 & 915 &   44.80  &  -14.20\\
A3560    &   1.721 & 5.03  &    -35.79 &   99.99 &   -1.57 &    0.04 &  -17.01 &   70.76 &   -4.67 &    7.80 &     0.51 & 5378 & 710 &   44.12  &  -14.26\\
A3667    &   3.603 & 0.74  &    -20.82 &    0.40 &   -0.65 &    0.14 &  -10.87 &    0.00 &    4.05 &    0.00 &    1.76 & 3451 & 993 &   44.94  &  -14.43\\
A3716    &   2.200 & 1.43  &    -22.41 &    0.70 &   -1.03 &    0.28 &  -16.77 &    0.78 &    0.17 &    1.09 &     0.83 & 3245 & 833 &   44.00  &  -14.01\\
A3809    &   1.298 & 1.77  &    -21.36 &    0.59 &   -1.15 &    0.16 &  -15.83 &    0.52 &   -3.80 &   11.62 &    1.22 & 1202 & 563 &   44.35  &  -14.71\\
A3880    &   2.358 & -1.71 &    -40.60 &   99.99 &   -1.13 &    0.62 &  -16.03 &    0.00 &  -21.69 &    0.00 &    2.73 & 2613 & 763 &   44.27  &  -14.48\\
A4059    &   2.142 & 2.47  &    -21.03 &    1.10 &   -0.77 &    0.66 &  -16.31 &    0.51 &    0.70 &    0.90 &    1.03 & 3527 & 715 &   44.49  &  -14.09\\
IIZW108  &   1.179 & 0.74  &    -21.61 &    0.56 &   -0.53 &    0.43 &  -18.62 &    0.32 &    0.12 &    0.42 &     0.20 & 1060 & 513 &   44.34  &  -14.28\\
MKW3s    &   1.116 & 2.04  &    -21.20 &    0.53 &   -0.98 &    0.10 &   -6.81 &    0.00 &   -2.94 &    0.00 &    1.97 & 983 & 539 &    44.43  &  -13.99\\
RX0058   &   1.652 & 3.32  &    -21.65 &    1.01 &   -1.17 &    0.17 &  -15.51 &    0.29 &   -2.32 &    0.59 &    1.60 & 2866 & 637 &   43.64  &  -14.29\\
RX1022   &   1.402 & 4.40  &    -20.97 &    1.07 &   -1.02 &    0.89 &  -17.41 &    1.78 &   -0.68 &    1.15 &    1.39 & 1309 & 577 &   43.54  &  -14.40\\
RX1740   &   1.436 & 4.59  &    -21.12 &    1.40 &   -1.27 &    0.20 &  -11.75 &   99.99 &    1.77 &   99.99 &    1.26 & 3153 & 582 &   43.70  &  -13.94\\
Z2844    &   1.199 & 5.54  &    -19.28 &    0.34 &    0.87 &    0.65 &  -17.17 &    0.57 &   -1.52 &    0.17 &    1.04 & 2168 & 536 &   43.76  &  -14.20\\
Z8338    &   2.148 & 3.04  &    -21.96 &    1.15 &   -1.19 &    0.16 &  -15.71 &    0.44 &    1.19 &    0.95 &    1.70 & 3910 & 712 &   43.90  &  -14.24\\
Z8852    &   2.124 & 3.20  &    -20.97 &    0.50 &   -1.10 &    0.17 &  -15.96 &    0.40 &    0.43 &    0.70 &    0.88 & 3463 & 765 &   43.97  &  -13.87\\
\hline
\multicolumn{16}{|c|}{Galaxies}\\
\hline
Median  &     -         & -             & -21.15 & -             & -0.97  & -            & -16.30 &-             & -0.60 & -         & -    & -    & -            & - & - \\
Mode     &     -         & -             & -21.75 & -             & -1.10   &  -           & -15.75 &-             & -1.70 & -       & -      & -    & -            & - & - \\
\hline
\multicolumn{15}{|c|}{Galaxies+Unknown}\\
\hline
Median  &     -         & -             & -21.45 & -             & -1.05   & -            & -16.24 &-             & -1.13 & -     & -        & -    & -            & - & - \\
Mode     &     -         & -             & -21.25 & -             & -1.10   &  -           & -16.25 &-             & -1.10 & -& -             & -    & -            & - & - \\
\hline
\multicolumn{16}{|c|}{With $\alpha_b=-1.1$ fixed}\\
\hline
\multicolumn{16}{|c|}{Galaxies}\\
\hline
Mode     &     -         & -             & -21.25 & -             & --   &  -           & -16.25 &-             & -1.50 & - & -            & -    & -            & - & - \\
\hline
\multicolumn{16}{|c|}{Galaxies+Unknown}\\
\hline
Mode     &     -         & -             & -21.25 & -             & --   &  -           & -16.25 &-             & -1.10 & -& -             & -    & -            & - & - \\
\hline
\end{tabular}
\end{center}
\label{tab:fitpar}
\end{table*}%

\section{Does the LF varies with cluster properties?}\label{sec:prop}
For the  WINGS clusters we possess two proxies of the global cluster mass, i.e. the velocity dispersion and the X--ray luminosity.
The first one has been calculated by \citet{Cava2009} using both our spectroscopic redshifts and the redshifts from the literature. We give here updated values that take into account the more recent data that have become available through the DR7 release of the SDSS spectroscopic survey \citep{dr7}.
We used as reference set of fitted parameters those found after having imposed the bright end slope ($\alpha_b=-1.10$), and then fitted a linear relation between the faint end slope and the cluster mass proxies (left panel of Fig.\ref{fig:lf_pars}), and between the bright end characteristic magnitude and the two proxies (right panel of Fig.\ref{fig:lf_pars}), in order to understand whether the cluster mass bears some influence on the final LF.
Superimposed to every plot are the linear relations, while the insets report the slope of the fitted linear relation together with the formal fit  1-$\sigma$ error. The shaded area is the RMS region derived considering a null variation of the faint end slope (left panel) and of the bright characteristic magnitude (right panel) with the mass proxies.

The only relation that appears significant is the one between the X--ray luminosity and the slope in the faint end (Fig.\ref{fig:lf_pars}, lower left panel), but even in this case the statistical analysis of the correlation using the Spearman/Kendall test gives a null correlation (the two correlation coefficients are -0.07 and -0.03, respectively, while their significance is 0.66 and 0.74).
Therefore, we conclude that for our sample of clusters we do not find any correlation of the LFs with the velocity dispersion and with the X--ray luminosity (i.e. with the mass) of the clusters.
We remember, though, that we are analyzing here galaxies located in the same physical region of the clusters (i.e. $0.5 \times r_{200}$), and we are using this population of galaxies to infer correlations with global properties of clusters. 
If differences from cluster to cluster arise in the external regions (as it seems the case, see \citealt{Hansen05,Popesso06}), they can be responsible for a different relation with the cluster global properties.

\subsection{The Dwarf--to--Giant ratio}
In order to verify our results, we decided to use a quantity not related to our fitting procedure,  being based only on galaxies number counts.
We then used the ratio between the number of faint galaxies and that of bright galaxies, the so called Dwarf--to--Giant ratio (DGR), to verify whether any relation exists between the overall description of the LF and the global cluster environment.
To be consistent in our definition of DGR, we counted dwarf galaxies only up to the brightest magnitude limit of the entire sample of clusters, i.e. $M_V=-15.15$.
To separate giant and dwarfs we used, instead, a value of $M_V=-19.0$.

\begin{figure}
   \centering
   \includegraphics[width=0.37\textwidth]{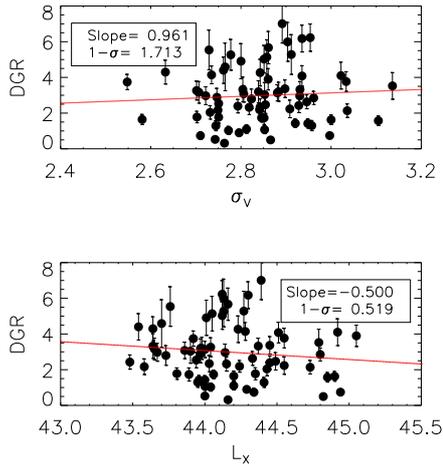}
      \caption{Dwarf--to--Giant ratio versus $\sigma_V$ (upper panel) and $L_X$ (lower panel), with superimposed the least square fit.}
         \label{fig:dgr}
   \end{figure}

We show in Fig.\ref{fig:dgr} how the DGR varies with the velocity dispersion (upper panel) and with the X--ray luminosity (lower panel).
Superimposed over both plots we also draw the least square fit to the data, that takes the errors into account.
The slope of the relation between DGR and $\sigma_V$ is $0.961$, with an error of $1.713$, and it is therefore compatible with being flat. On the other hand, the relation with the X--ray luminosity has a slope of $-0.500 \pm 0.519$. RX1740 has been excluded from the plot to better visualize the data, but it is included in the fit.

Again, there is a hint for less massive clusters (as traced by their X--ray luminosity) hosting a larger number of dwarf galaxies with respect to massive galaxies, but only if excluding the clusters where the DGR shows larger errors. If including the whole cluster sample, instead, both the relation with the X-ray luminosity and the one with the velocity dispersion are not significant.
In fact, the Spearman correlation test confirms that there is no correlation at all between the DGR ratio and the mass of the cluster.

\subsection{LF of different subsamples}
Here we consider various sub-samples of clusters for which we construct the LF.  First we analyzed the LF of two subsamples characterized by extreme values of X-ray luminosity and velocity dispersion. We select the 10 clusters with the highest (lowest) X-ray luminosity and the 10 clusters with the highest (lowest) velocity dispersion.

   \begin{figure}
   \centering
   \includegraphics[width=0.33\textwidth,angle=90]{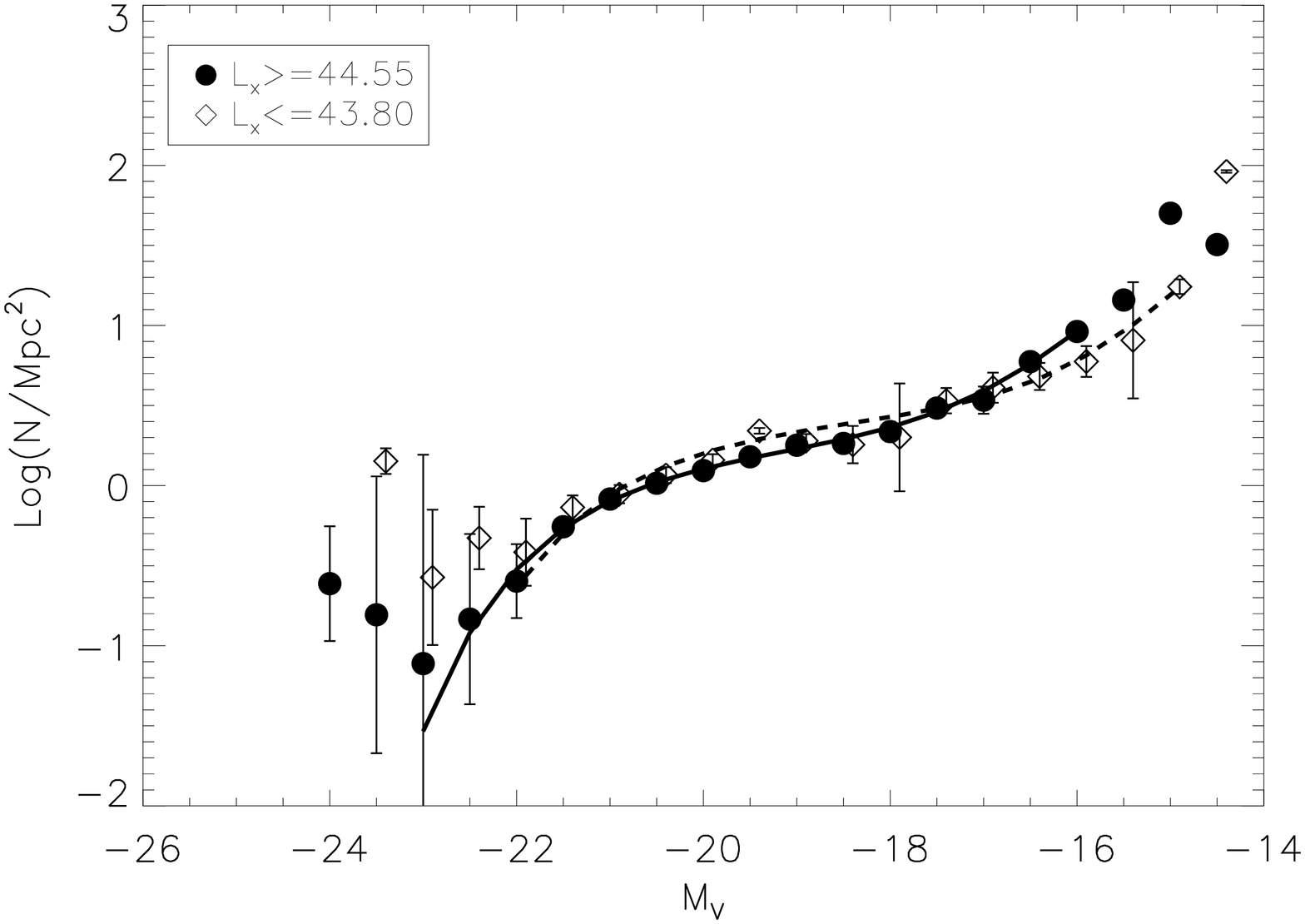}
   \includegraphics[width=0.33\textwidth,angle=90]{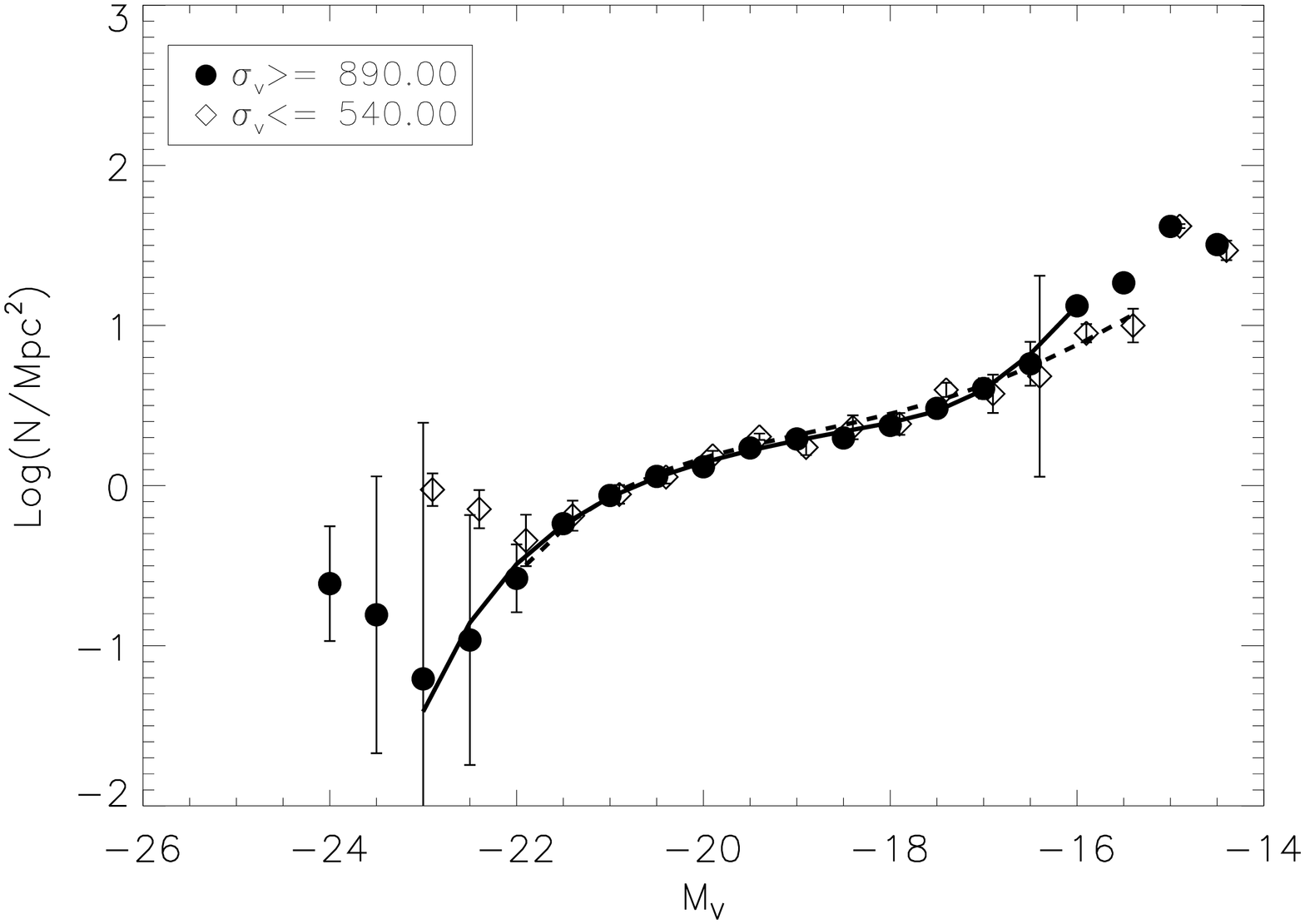}
      \caption{Composite Luminosity Function of galaxies belonging to the 10 clusters with highest (and lowest) X-rays luminosities in the upper panel, and to the 10 clusters with the highest (and lowest) velocity dispersions in the lower panel. The fits are drawn with a continuous line for the highest X--ray luminosity (velocity dispersion) samples, and with a dashed line for the lowest X--ray luminosity (velocity dispersion).}
         \label{fig:lfcomp_lxsig}
   \end{figure}

In Fig.\ref{fig:lfcomp_lxsig} we show the composite LF (i.e. the LF obtained from the single LFs by summing all contributions after having normalized them to have the same number of objects above a certain magnitude, see Sec. \ref{sec:composite_lf} for our own definition of composite LF) for these subsamples of clusters: in the upper panels WINGS clusters are subdivided according to their X--rays luminosity (taken from \citealt{Ebeling1996,Ebeling1998,Ebeling2000}), while in the lower panels they are separated on the basis of their velocity dispersion. In both figures the filled symbols represent the LF for the sample with highest X--rays luminosity (velocity dispersion), while open symbols refer to the ones with the lowest X--rays luminosity (velocity dispersion). 
In order to better compare the two samples they have been normalized so that they possess the same number of galaxies brighter than $M_V=-19$.
Superimposed are the two fits (in continuous and dashed, respectively, for the two subsamples) that we obtained leaving free all parameters of the double Schechter function. In order not to be biased by low statistics, we fit only points where the global contribution comes at least from 5 clusters. For this reason the bins brighter than $M_V=-23$ never contribute to the fit, and clusters with the smallest values of $L_X$ and $\sigma_V$ have been fitted up to $M_V=-22$. 

The central part of the LF is very similar in the two subsamples, indicating that differences, if any, arise at the two extremes of the distributions.
 More massive clusters have a brighter characteristic magnitude and a steeper slope in the bright regime, with respect to clusters with smaller masses (see \citet{Hansen05} and  \citet{croton+2005} for similar results derived from local densities).
The slope in the faint end part of the LF is $-2.4\pm 0.1$ and $-2.5 \pm 0.4$ in the two subsamples, respectively, when looking at the trends with the X--ray luminosity, while it turns out to be $-2.6 \pm 0.4$ and $-2.1 \pm 0.3$, respectively, when dividing the samples according to their velocity dispersion.
Therefore, given the uncertainties, we can conclude that the two shapes of the LF are very similar in both cases, confirming the results found in previous sections.

\section{Composite Luminosity Function}\label{sec:composite_lf}
The lack of any significant relation between the single cluster LFs and the overall cluster properties, led us to put more stringent constraint on our result by constructing the so called composite LF.
We calculated it by summing all the clusters LF after having normalized them in order to have the same number of objects brighter than $M_V=-19$. To construct the LF we follow a modified version of the formulation given by \citep{Colless89,Popesso06}. The number of galaxies $N_j$ in the final LF in the absolute magnitude j-th bin is therefore calculated as:

\begin{equation}
N_j=\frac{N_{c,0}}{m_j^2}\times\sum_{0}^{m_j}{\frac{N_{i,j}}{N_{i,0}}}
\end{equation}

where $N_{c,0}$ is the total number of galaxies brighter than $M_V=-19$, $m_j$ is the number of clusters contributing to the j-th bin, $N_{i,j}$ is the number of galaxies in the j-th bin coming from the i-th cluster and $N_{i,0}$ is the number of galaxies in the i-th cluster brighter than $M_V=-19$. Here we use $m_j^2$ instead of  $m_j$, as in the original formalism by \citet{Colless89}, in order to end up with a LF representative of the average cluster. In fact if we suppose to have an ideal situation of $m_j$=n identical clusters with $N_{i0}$=$N_{norm}$ and $N_{i,j}$=$N_j$ then:

\begin{equation}
N_{c,0}= \sum_{i}^{n}{N_{i,0}}=n\times N_{norm}
\end{equation} 

using the original formalism of \citet{Popesso06} and substituting equation (6) in it we can see, after simple algebra, that $N_{cj}$=$N_j\times n$; therefore in the original form, the LF results in {\it n} times the single LF which is not a "true" LF. We avoid this by dividing the original expression by the factor $m_j$ which is the number of clusters used in each bin. The errors on the single bin are derived as the squared root of the sum of the single variances divided by the number of clusters contributing to the given bin.

   \begin{figure}
   \centering
   \includegraphics[width=0.37\textwidth,angle=90]{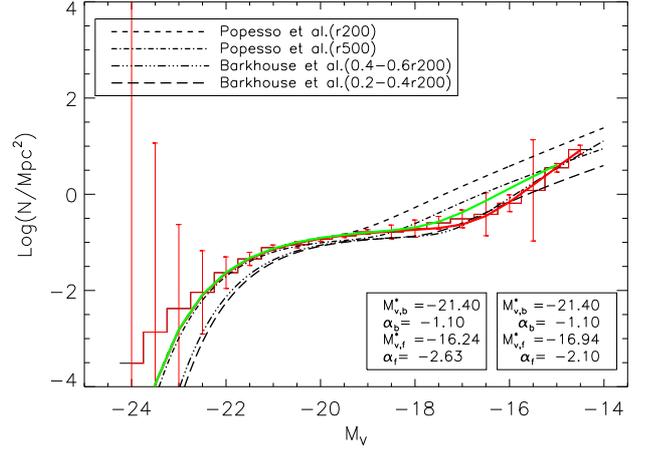}
      \caption{Composite Luminosity Function of WINGS galaxies. Superimposed are the double Schechter fits obtained having imposed the bright end slope $\alpha_f=-1.10$: red for the population of galaxies, green for the population of galaxies and unknown. The two insets in the lower right corner are the values of the fit. The black lines are fits taken from the literature (see the top left inset). }
         \label{fig:lf1scomp}
   \end{figure}
   
   Fig.\ref{fig:lf1scomp} shows the derived distribution for the sample of WINGS galaxies. In red and green  are shown the two fits obtained to the sample of galaxies (plus galaxies unknown) and to the secondary sample of all objects (i.e. galaxies and unknown, without the sources classified as stars). While in the bright part the two fits are coincident, soon after the central plateau the mixed distribution starts rising while the pure galaxy population remains flatter. In particular in the bright part the LF is well constrained, and it does not depend on the objects classification.  At low luminosities where the classification of objects becomes more difficult, i. e. the galaxies/unknown separation is a critical issue, the LF varies in the two subsamples (as expected).
   
In particular, the best fit to our LF, after having fixed the bright end slope, as we did for the single luminosity functions, gives $M_{V,b}=-21.40$, $M_{V,f}=-16.24$, and $\alpha_f=-2.63$ when we consider the galaxy population, while in the faint end we find $M_{V,f}=-16.94$, and $\alpha_f=-2.10$ when including the unknown objects. 
In the plot we also superimpose the LF fit derived by \citet{Popesso06} and \citet{Barkhouse07}, rescaled to match the bright part of the LF.

When we consider the sample of galaxies (plus galaxies unknown) our LF (and consequently its fit) is slightly different from the ones given in literature, even if still compatible. We find a steeper rising in the faint end regime and a more pronounced central plateau. 
If we include in our sample all unknown objects, instead, we find a  better agreement, with a flatter slope and a brighter characteristic magnitude in the faint end part of the LF. However, the literature fits present a still higher number of low luminosity objects, probably suggesting that the contribution of spurious classifications in the SDSS samples is not negligible and that our WINGS sample of galaxies (plus unknown galaxies) is a good tracer of the population of cluster galaxies at least in the range of magnitudes we are using.

\subsection{LF of galaxies with different morphologies}\label{sec:LF_morph}

   \begin{figure}
   \centering
   \includegraphics[width=0.33\textwidth,angle=90]{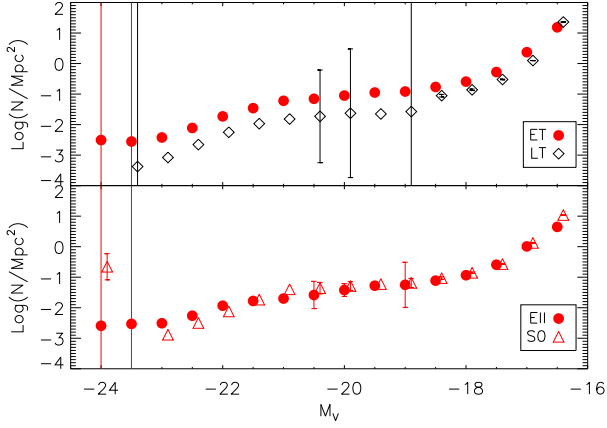}
      \caption{Composite Luminosity Function of galaxies classified as early--type (purple dots) and late--type (black diamonds) in the upper panel, while in the lower panel there are ellipticals (red dots) and S0 (red triangles).}
         \label{fig:lfcomp_morph}
   \end{figure}

 One of the main characteristics of clusters is the morphological mix of its galaxies. To test any dependence of the LF on morphology we constructed the LF of galaxies according to their morphology. 

Fig.\ref{fig:lfcomp_morph} shows the LFs of early--type galaxies (in red dots) and late--type galaxies (in black diamonds). In the bottom panel we show the LF for ellipticals (red dots) and S0 (red triangles) galaxies. 
As already noted in \citet{Vulcani+2011b}, the population of early--type galaxies is always predominant over the contribution of the late--type ones.
The two shapes of the LF are different, with the late--type galaxies showing a more flat central plateau, and a rapid decline at both bright and faint luminosities.
As for the contribution of ellipticals and S0s, we show in Fig.\ref{fig:lfcomp_morph} (lower panel) that the two populations have almost the same trend along the whole LF. However, at bright luminosities ellipticals outnumber S0s, while in the central plateau S0s seem to give a larger contribution (see also \citealt{Vulcani+2011b} for the same conclusions about the mass functions).

This trend is partially at odds with what found by \citet{Popesso06}, that evidence a predominant fraction of early--types in the faint end regime. However, their classification was mainly based on galaxy colors.

The analysis by \citet{Popesso06} shows also a predominance of late--type dwarfs when moving towards the external regions of clusters. Unfortunately we can not confirm yet this result at our redshifts, where clusters need larger CCDs to reach the $r_{200}$ limit. We remind, however, that our classification is based on morphological criteria, and not on the galaxy colors.

\section{Spectroscopic Luminosity Function}\label{sec:spec_lf}
Given the high spectroscopic coverage of our cluster sample, we finally calculated the spectroscopic LF.
As previously done in other works \citep{Vulcani+2011b}, we decided here to consider only those clusters that have a spectroscopic completeness larger than 50\%, i.e. 21 out of 48 clusters. 
The sample is made of 2009 galaxies that are cluster members.

   \begin{figure*}
   \centering
   \includegraphics[width=0.63\textwidth,angle=90]{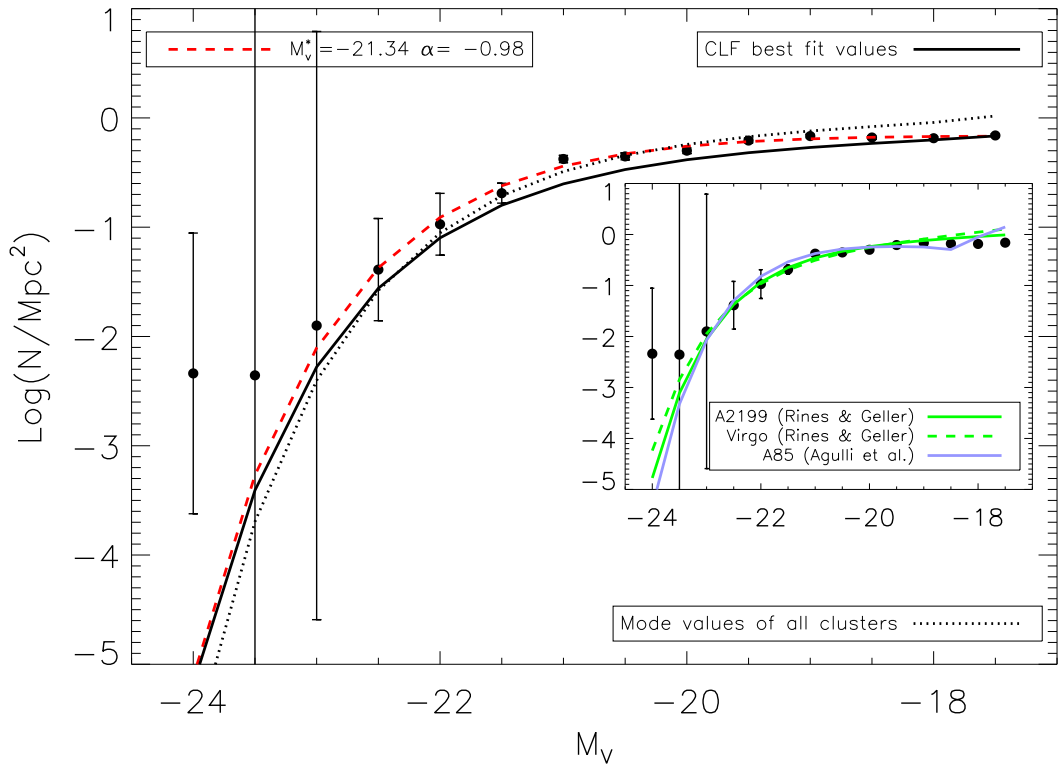}
      \caption{Spectroscopic Luminosity Function of WINGS cluster members. The single Schechter fit to the SLF is drawn in red while in black are superimposed the fit to the photometric LF (continuous line), and the Schechter function obtained using the median values of the whole sample (dotted line). In the inset we compare our SLF with the ones by \citet{rines+geller2008} for Virgo and A2199, and by \citet{agulli14} for A85.}
         \label{fig:lfcomp_spe}
   \end{figure*}

Fig.\ref{fig:lfcomp_spe} shows the spectroscopic LF (SLF) of the sample, after the correction made using both the spectroscopic and the photometric completeness, as described in \citet{moretti14}, sec.6.3.
The  LF is obviously less deep than the photometric one, and we decided to keep only points reaching -17.5 in V band absolute magnitude, as in this magnitude bin 17/21 clusters contribute to the galaxy population. 
The best--fit of a single--Schechter function is superimposed to the points we draw in dashed red (the values of the fit are given in the top left label). We show also the best fit of the photometric LF, normalized to the same constant (dotted line) and the fit obtained imposing the mode values of the entire sample (continuous line).

The SLF fit is very similar to the best fit of the photometric CLF, being $M_V^\ast = -21.34 (-21.40) \pm 0.17 (0.07)$ for the two fits, respectively. The mode of the global LF, when considering the population of galaxies corrected for the fraction of unknown that could be considered as galaxies, is $M_V^\ast=-21.25 \pm 0.15$.
Our spectroscopic and photometric LFs are therefore in agreement, even if this can be probed only in the bright part of the LF.

Previous studies based on samples of spectroscopically confirmed cluster members found slightly steeper slopes for the Schechter fit: \citet{DePropris03} analyzing 2dFGRS data in the {\it b}$_J$ band gives $\alpha=-1.28$, while \citet{Christlein03} using R band data in 6 clusters estimated $\alpha=-1.21$.
 Their magnitude limits extends from 3 to 7 magnitudes below $M^\ast$, and it is therefore only marginally comparable with our observational range. In fact, their slope is derived using magnitudes where we do expect to find an upturn, but it is less pronounced than the one we find using the photometric sample.
It is interesting the comparison of our SLF with the results found in the recent works by \citet{rines+geller2008} and \citet{agulli14} in which they derive the spectroscopic luminosity function for Abell 2199, Virgo and Abell 85, respectively,  reaching very faint magnitudes. In particular there is a good agreement of our SLF with the corresponding bright part of all SLFs, i.e. in the region where we possess spectroscopic information. In the faint end regime we find a slightly shallower slope (-0.98 to be compared with -1.28 and -1.13 for Virgo and A2199, and with -1.58 found in A85).

\section{Conclusions}\label{sec:conclusions}

We studied the LF in a sample of WINGS clusters up to $0.5 \times r_{200}$. This allows us to evaluate the cluster LF  in the same physical region in terms of radial coverage. After a careful field subtraction using the work of \citet{Berta06} that has been obtained with our own observational setup we cleaned the samples from stars and background detections and we find that a fit with a single Schechter function is not able to reproduce the entire range of luminosity distribution, and we therefore moved to the widely used approach of fitting a double Schechter function to our LFs.

First, we addressed the still unsolved question regarding the universality of the LF.
 We find that a large spread exist among values for single clusters, and the agreement with other studies is satisfying only when comparing the bright part of the LF, while in the faint end discrepancies arise.
The fitted values for a single cluster do not depend on the global characteristics of the cluster itself, such as the X--ray luminosity or cluster velocity dispersion, which are, however, quantities derived for the global cluster (while the LF covers only the internal region).
We find that in the LFs for extreme subsamples of galaxies, i.e. those showing the highest (and lowest) values of $L_X$ and $\sigma_V$, the overall shape of the two distributions is preserved. The DGR as well does not depend on cluster's masses (as derived from the same proxies).

We constructed the composite luminosity function, by stacking all the LFs. This approach compensated by a larger statistics the errors on single cluster LF.
This LF is in excellent agreement with what previously found by other studies in the bright part of the LF, while it shows a slightly steeper trend in the faint region. This steeper slope is somewhat compensated by the presence of a fainter characteristic magnitude, which leads to a more pronounced central plateau.
If we include in the data detections belonging to the unknown class we recover the faint end slope. We conclude that a careful object classification, possible only in dedicated survey such as ours, is the only way we have to discriminate which one of the two slopes is more probable. We point out in addition that our LF  is in good agreement with the recent findings by \citet{agulli14} that derived a spectroscopic luminosity function down to $M_r\sim$-16.0 for a cluster belonging to our sample.

We also used the morphological classification given by MORPHOT  to derive the LFs of galaxies with different morphologies, up to the limit where the morphological classification was available for at least half of the photometric sample. We find that early--type galaxies dominate the LF over the entire magnitude range.
Among early--type galaxies we find that ellipticals slightly outnumber S0s in the bright end, while the S0 fraction seems to increase in the central plateau.

Finally, we used a restricted sample of galaxies for which we have the spectroscopic membership confirmation to derive a clean LF. We obtained the LF only for the 21 clusters where the spectroscopic completeness is larger than 50\%. Even though this LF is not as deep as the photometric one (as expected), in the bright end we can confirm the values that we found from the photometric LF.

Our study indicates that the faint end LF slope might have been overestimated in the past, thus leading to a LF steeper than the real one. This aspect needs to be assessed, in order to link the presence of dwarfs to the cosmological predictions and/or to the higher redshift results (where, though, their mere presence is still debated, see \citealt{harsono+depropris2009} and \citealt{crawford+2009} for different results).
They seem to be equally divided into early and late morphologies, and among the early types they are again equally divided into Ellipticals and S0s. When looking at the overall composition of the LF, instead, we find mainly S0 in the central plateau, and mainly Ellipticals in the bright part of the LF. This could be the sign that between high and low redshift small S0s form by merging of small late-types.
However, deeper studies of local clusters by \citet{trentham+tully2002,Hilker+03,misgeld+2009} have demonstrated that the faint LF is much flatter than what emerges from pure photometric studies (even if they looked for early--type galaxies), thus posing a dramatic challenge to the theoretical predictions by \citet{moore+1999,jenkins+2001} of a steep slope $\alpha=-2.0$.

\begin{acknowledgements}
We acknowledge the financial contribution from PRIN/INAF2014 "Galaxy evolution from cluster cores to filaments". BV ackowledges the World Premier International Research Center Initiative (WPI), MEXT, Japan and the Kakenhi Grant-in-Aid for Young Scientists (B)(26870140) from the Japan Society for the Promotion of Science (JSPS). We thank the anonymous Referee whose comments greatly improved the paper.
\end{acknowledgements}

\bibliographystyle{aa} 
\bibliography{wings_lf.bib} 

\end{document}